\documentclass[%
 reprint,
 amsmath,amssymb,
 aps,
]{revtex4-1}
\usepackage{graphicx}
\usepackage{dcolumn}
\usepackage{bm}
\usepackage{gensymb}
\usepackage{hyperref}
\usepackage{longtable}
\usepackage{booktabs,array}
\usepackage{hhline}

\begin{document}

\preprint{APS/123-QED}

\title{Determination of photonuclear cross section of $^{61}$Ni($\gamma$,xp) reaction via surrogate ratio technique}

\author{Shaima Akbar$^1$}\email{shaimaakbar333@gmail.com}
\author{M.M Musthafa$^1$}\thanks{mmm@uoc.ac.in}%
\author{C.V Midhun$^1$}
\author{S.V Suryanarayana$^2$ $^3$}
\author{Jyoti Pandey$^4$ $^5$}
\author{Bhawna Pandey$^4$}
\author{A Pal$^2$}
\author{P.C Rout$^2$}
\author{S Santra$^2$}
\author{Antony Joseph$^1$}
\author{K.C Jagadeesan$^6$} 
\author{S. Ganesan$^7$} 
\author{H. M. Agrawal$^4$}

\affiliation{$^1$ Department of Physics,University of Calicut, Calicut University P.O Kerala, 673635 India}
\affiliation{$^2$ Nuclear Physics Division, Bhabha Atomic Research Centre, Mumbai 400085, India}
\affiliation{$^3$Manipal Centre for Natural Sciences, MAHE, Manipal - 576014, India}
\affiliation{$^4$  Department of Physics, G.B. Pant University of Agriculture and Technology Pantnagar, Uttarakhand 263 145, India}
\affiliation{$^5$Inter University Accelerator Centre, New Delhi, Delhi 110067, India}
\affiliation{$^6$ Radiopharmaceuticals Division, Bhabha Atomic Research Centre, Mumbai 400085, India}
\affiliation{$^7$ Formarly Raja Ramanna Fellow, Bhabha Atomic Research Centre, Mumbai 400085, India}

\date{\today}

\begin{abstract}
The photo nuclear reaction cross section of $^{61}$Ni($\gamma$,xp) reaction have been measured by employing surrogate reaction technique. This indirect method is used for the first time to obtain the cross section of photo nuclear reaction. The compound nucleus $^{61}$Ni$^{*}$ was populated using the transfer reaction $^{59}$Co($^{6}$Li,$\alpha$) at E$_{lab}=$ 40.5 MeV. To calculate the surrogate ratio,  $^{60}$Ni($\gamma$,xp) was selected as reference reaction and the corresponding compound nucleus $^{60}$Ni$^{*}$ was populated using the transfer reaction $^{56}$Fe($^{6}$Li,d) at E$_{lab}=$ 35.9 MeV. The experimental cross section data of  the reference reaction has been taken from EXFOR data libraries. Compound nuclear cross section calculations have been done using EMPIRE 3.2.3 code.
\end{abstract}

\maketitle


\section{\label{sec:level1}introduction}
The photonuclear reaction cross section data have a significant impact on several fields of science and technology like dose calculation in radio therapy, nuclear astrophysics, analysis of radiation transport, designing radiation shield, nuclear waste transmutation and activation analysis for material  identification.  It is also relevant in the study of influence of photoreactions on neutron balance inside fission reactors and to evaluate the damages in structural materials of reactors. Presently due to the lack of evaluated data, users depend on raw photo-nuclear data or primary cross sections from different measurements\cite{iaea}. The energy dependence of cross section, energy of dipole state, giant resonance width, cross section magnitude\cite{william} can be determined from photo-nuclear cross section measurements.
\paragraph*{}In the early works on photon induced nuclear reactions, $\gamma$-rays from neutron or proton induced reactions are used as photon sources\cite{kats, evans}. The $^{7}$Li(p,$\gamma$) reaction is the main source of 17.6 MeV $\gamma$-rays which itself is a broad spectrum. Further in such cases the energies available are very limited. As an alternate method, the bremsstrahlung photons produced using high-Z target materials is a common source of photons for photon induced nuclear reactions\cite{Berman}. Such photons are easier to produce via electron accelerators. Bremsstrahlung photons are the continuous photon sources and requires unfolding of the bremsstrahlung spectra to generate the required cross section\cite{penfold}. Recently Midhun et al\citep{midhun} measured medium energy bremsstrahlung beam produced from a medical linear accelerator by detecting recoiled electrons corresponding to 180 deg Compton scattering. Quasimonoenergetic photons, available from in-flight positron annihilation, is also used as photon sources. However, the intensity of such photon source is very low. Further some sort of systematic disagreements  are there while comparing the results obtained in bremsstrahlung experiments and the quasimonoenergetic annihilation experiments. Other experimental methods such as Laser Compton scattered photons, electron tagged bremsstrahlung photons etc are also used to generate the energy specific cross sections\cite{zen, haya}. These are odd facilities with limited access. In this scenario, surrogate reaction techniques\cite{escher}, which is an indirect method, to be considered as better solution to overcome these kind of limitations. 
\paragraph*{} Surrogate reaction technique is developed on the basis of Bohr's hypothesis, which states that the formation and the decay of compound nucleus are independent of each other. Therefore same compound nuclear state of interest, in the desired reaction, can be populated through different entrance channels and the decay probabilities corresponding to compound nuclear state can be studied, hence it is a two step process. Surrogate method is commonly used for the determination of cross section of certain reactions in which either the particle beam or target is not readily available. In the surrogate reaction method, a single beam energy for the projectile in the transfer reaction, is enough to produce the compound nucleus of interest, over wide range of excitation energies. This in turn can allow the determination of cross section of required reaction over a range of excitation energies. The cross section can be calculated by applying Weisskopf-Ewing limit of the Hauser-Feshbach theory. Surrogate technique was first used in determining neutron induced fission (n,f) cross sections\cite{cramer}. Later a series of experiments were carried out for studying neutron capture (n,$\gamma$) cross sections, neutron induced fission cross sections and neutron induced charged particle emissions\cite{brit, escher2, kesse, jp, bp, raman}. The intricacy in determining efficiency is the major limitations in determining absolute surrogate method. To overcome these limitations the surrogate ratio approach was developed\cite{nayak, asim, escher}.
In a photonuclear reaction, involving medium mass and heavy mass nuclei the target nucleus may absorb photon and reaches a highly excited intermediate state, called dipole state. This dipole state proceeds similar to compound nuclear evaporation, where the excited nucleons undergo many interactions, before escaping the nuclear surface. Thus the excited nucleus no longer have the memory of the absorption process. This is the key point for considering  evaluation of cross section of photonuclear reactions using surrogate ratio method. 
\paragraph*{}In the present work, cross sections for $^{61}$Ni($\gamma$,xp) reactions have been determined via surrogate technique, employing surrogate ratio method. The compound nucleus $^{61}$Ni$^{*}$ was populated through the transfer reaction $^{59}$Co($^{6}$Li,$\alpha$) at E$_{lab}=$ 40.5 MeV. The compound nucleus is formed over wide range of excitation energies with $^{6}$Li beam energy 40.5 MeV. The experimental cross section of $^{60}$Ni($\gamma$,xp) reaction, taken from EXFOR data library is used as the reference reaction data in order to determine the $^{61}$Ni($\gamma$,xp) cross section by applying surrogate ratio method. As of our knowledge cross section for this reaction is being reported for the first time. Nickel is an important structural material in fusion and fission reactors. Though the abundance of  $^{61}$Ni is (1.14\%) very low, this isotope is produced in large quantities through neutron induced channels from other highly abundant isotopes over a long duration of exposure.  This $^{61}$Ni may be exposed to high energy prompt gamma rays and that can lead to excess proton production which, in turn, can cause damages to the structural materials. Further in the case of fusion reactors this will induce a serious problem on account of difference in magnetic properties of the nickel in the core magnet and the cobalt formed in the ($\gamma$,p) reactions. This can affect the plasma stability.

 \paragraph*{}
\section{Experimental details}
 The experiment was performed at the BARC-TIFR Pelletron Accelerator Facility, Mumbai, India. $^{6}$Li beam obtained from the palletron was used as the projectile for both the transfer reactions. The self supporting thin metallic targets of natural cobalt ($^{59}$Co is 100\% abundant) and iron ($^{56}$Fe is 92\% abundant) were prepared by vacuum evaporation technique and rolling method respectively, with thickness $\approx$ 700 $\mu$g/cm$^{2}$. The compound nucleus $^{61}$Ni$^{*}$ and $^{60}$Ni$^{*}$ were populated at same excitation energies 
$\approx$ 21-32 MeV with beam energy of 40.5 MeV incident on $^{59}$Co and 35.9 MeV on $^{nat}$Fe respectively. The surrogate reactions for the desired and reference photon induced reactions, their ground state Q values (Q$_{gg}$ ) and the corresponding compound nucleus of interest are tabulated in Table \ref{Tab:1}.
 \begin{table}
\caption{Photon induced reactions and corresponding surrogate reactions in the present experiment, their ground-state Q values (Q$_{gg}$ ) and the compound nucleus
(CN) formed}
\begin{tabular}{c  c  c  c  c }
\hline \hline
E$^{^{6}Li}_{beam}$ \hspace{1mm} &  Surrogate \hspace{1mm} & Q$_{gg}$ \hspace{1mm}&  CN \hspace{1mm} &  Equivalent Photon\hspace{1mm} \\
(MeV)\hspace{1mm} & reaction \hspace{1mm} & (MeV) \hspace{1mm} & \hspace{1mm} & induced reaction \hspace{1mm}\\
\hline
\vspace{2mm} 
40.5 \hspace{1mm}& $^{59}$Co($^{6}$Li,$\alpha$) \hspace{1mm}& 13.65 \hspace{1mm}& $^{61}$Ni$^{*}$ \hspace{1mm}& $^{61}$Ni($\gamma$,xp)\hspace{1mm}\\ \vspace{1mm}
35.9 \hspace{1mm}&$^{56}$Fe($^{6}$Li,d) \hspace{1mm}& 4.817 \hspace{1mm} & $^{60}$Ni$^{*}$ \hspace{1mm} &  $^{60}$Ni($\gamma$,xp)\hspace{1mm} \\

\hline\hline
\end{tabular}
\label{Tab:1}
\end{table} 
\begin{figure}[h]
\includegraphics[scale=.3]{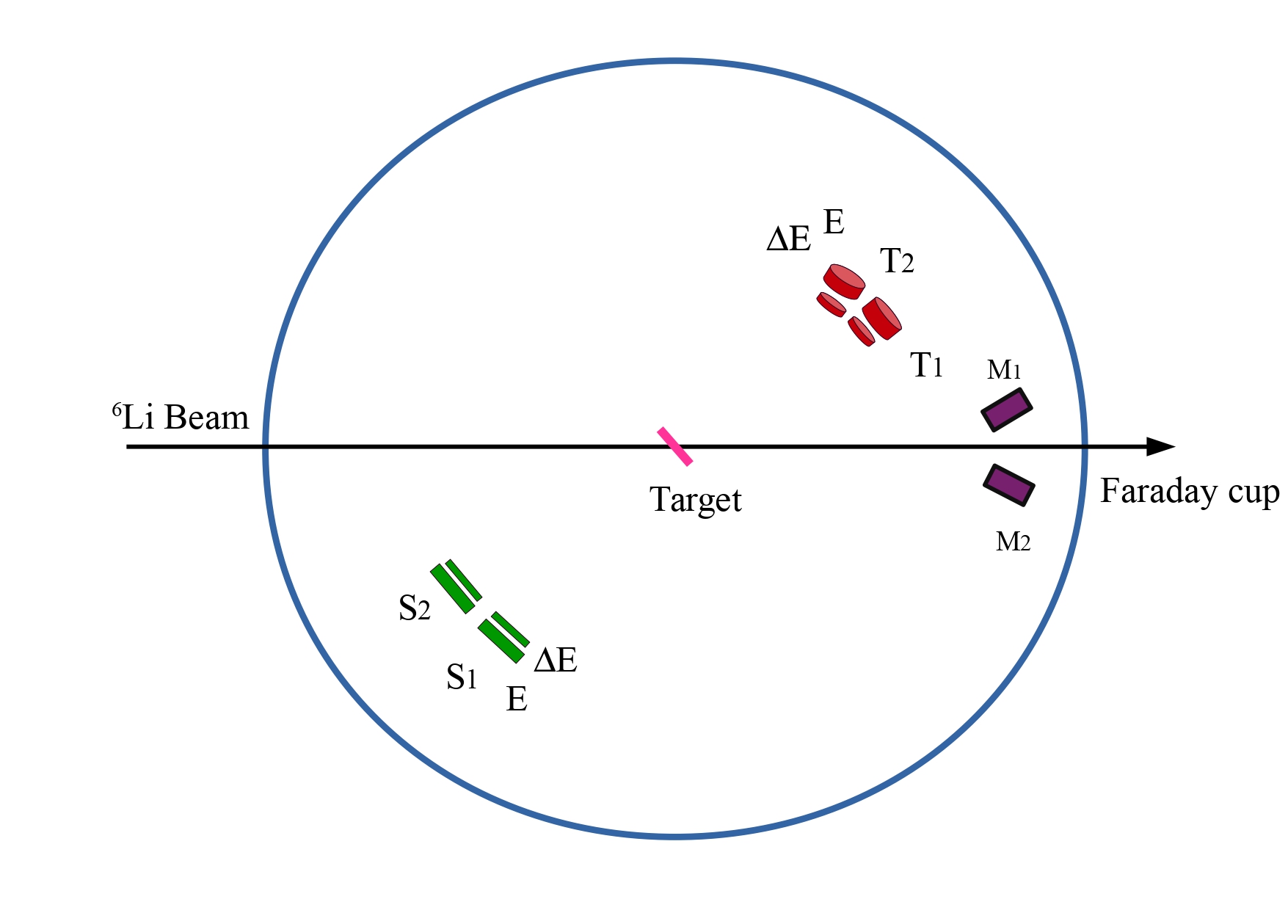}
\caption{\label{fig:one}The schematic diagram of experimental set up. T$_{1}$ and T$_{2}$ represents the particle telescopes for identifying PLFs from the transfer reactions and S$_{1}$ and S$_{2}$ are the strip detectors to detect decaying particles from the compound nuclei. The detectors are placed inside the scattering chamber of 1.5 m diameter.}
\end{figure} 
 \paragraph*{}To identify the projectile like fragments(PLFs), two silicon surface barrier detector telescopes T$_{1}$ and T$_{2}$ were mounted inside the scattering chamber at an angle of 25$\degree$ and 35$\degree$ from the beam direction around the grazing angle. Telescope T$_{1}$ consists of $\Delta$E-E detectors with thickness of 150 $\mu$m and 1 mm and T$_{2}$ consists of $\Delta$E-E detectors with thickness of 100 $\mu$m and 1mm respectively. To record the evaporated particles from the compound nucleus, 2 sets of strip detectors S$_{1}$ and S$_{2}$ are placed in the backward angles covering an angular range of 110$\degree$-130$\degree$ and 140$\degree$-160$\degree$. Thickness of $\Delta$E strip detector is $\approx$ 60 $\mu$m and that of E is $\approx$ 1500 $\mu$m and having an active area $\approx$50 mm$\times$50 mm. Experimental set up for the present measurement is shown in Fig.~\ref{fig:one}. The evaporating protons from the compound nucleus $^{61}$Ni and $^{60}$Ni in the two different transfer reactions are identified in strip detectors in coincidence with the outgoing deuteron for $^{60}$Ni and outgoing alphas for $^{61}$Ni using the same telescope. Correlation plot of $\Delta$E-E detectors clearly identifies each projectile like particles, which is shown in Fig.~\ref{eq:two}. 
  In order to isolate the random coincidences a TAC between telescopes and strips were generated. This ensures the PLF and evaporated particles are corresponding to the same compound nucleus.
  
\begin{figure}[h]
\includegraphics[scale=.30]{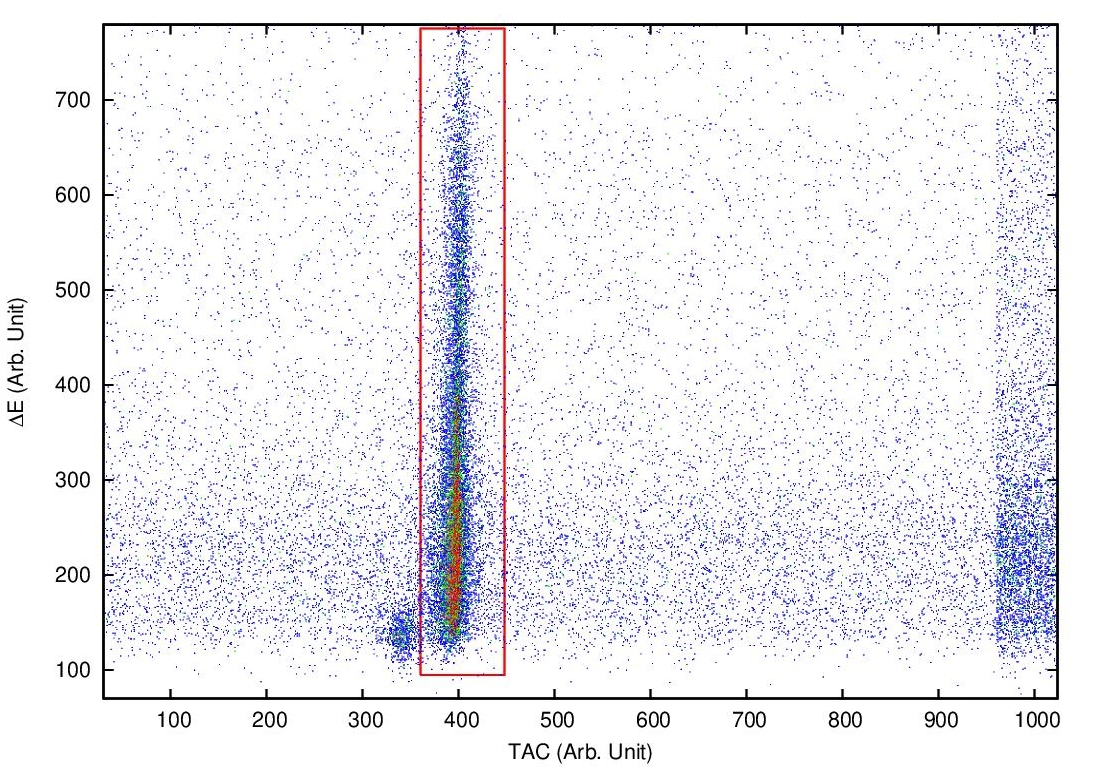}
\caption{\label{fig:three} Plot of proton TAC versus alpha PLF energy in $^{59}$Co($^{6}$Li,$\alpha$) reaction at E$_{lab}$ = 40.5 MeV}
\end{figure}

  A two dimensional plot of TAC versus $\Delta$E, including the gate for coincidence event are  shown in fig.~(\ref{fig:three}).
   The telescopes T$_{1}$ and T$_{2}$ were calibrated using the deuteron energy spectrum of $^{16}$O$^{*}$ obtained from the experiment $^{12}$C($^{6}$Li,d)$^{16}$O$^{*}$ at 18 MeV. Strip telescopes S$_{1}$ and S$_{2}$ were calibrated using the known energies of $\alpha$ particles from Pu-Am $\alpha$ sources.  
 \begin{figure}[h]
\includegraphics[scale=.6]{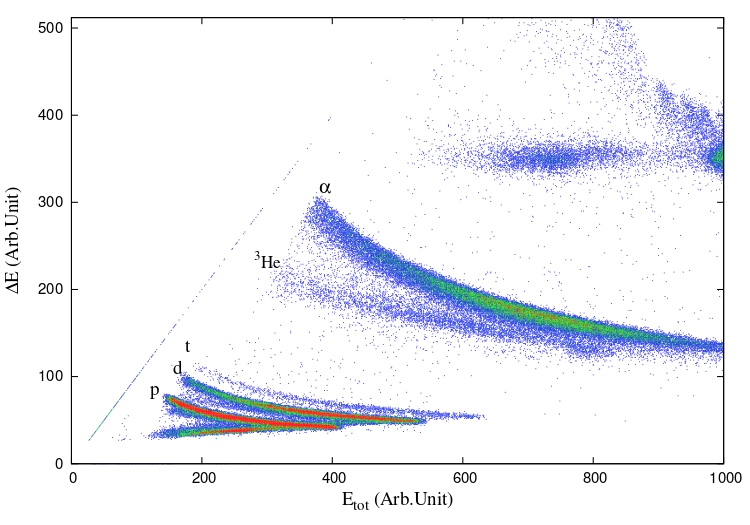}
\caption{\label{fig:two} E-$\Delta$E versus total energy E$_{tot}$ plot for $^{59}$Co($^{6}$Li,$\alpha$) reaction at E$_{lab}$ = 40.5 MeV measured at T$_{2}$}
\end{figure}
\section{data analysis}
\paragraph*{}  In the present study, the compound nuclei are formed in the excitation energy range of $\approx$ 21 - 30 MeV corresponding to the alpha energies, E$_{\alpha}=$ 35.5 to 23.5 MeV. The excitation energies are determined from PLF's using two body kinematics, applying appropriate gate to the PLF band. The equivalent photon energy range obtained is 21 - 30 MeV using the general expression $E_{a}= \frac{A_{a}+A_{A}}{A_{A}}(E_{ex}-S_{a})$ where $S_{a}$ is the projectile separation energy from the compound nucleus. Since we are dealing with the energetic gamma photon, the excitation energy will be equivalent to the incident energy after subtracting necessary recoiling energy of the nucleus.
\paragraph*{} The decay probability of the populated compound nucleus for the particular emission channels(here proton) at given excitation energy E$_{ex}$, $\Gamma_{p}^{CN}(E_{ex})$ is determined using the following relation;
\begin{equation}\label{eq:one}
\Gamma_{p}^{CN}(E_{ex})=\frac{1}{\epsilon}\frac{N_{i-p}(E_{ex})}{N_i(E_{ex})}
\end{equation}
Where N$_{i}$ represents the total number of PLF events and N$_{i-p}$ is the number of coincidence events, between the protons and PLF's, 
 $\epsilon$ denote the efficiency of detection set up.

\paragraph*{} Using the coincidence and single counts obtained from the telescopes, the decay probability of compound nucleus $^{61}$Ni in deuteron transfer reaction and $^{60}$Ni in alpha transfer reaction have been determined for excitation energies of interest in respective cases, by applying Eq.~(\ref{eq:one}) in steps of 1 MeV. 
 Accordingly, the cross section of $^{61}$Ni($\gamma$,xp) reaction is expressed in terms of surrogate ratio, as
\begin{equation} \label{eq:two}
\frac{\sigma^{^{61}Ni(\gamma,xp)(E_{\gamma})}}{\sigma^{^{60}Ni(\gamma,xp)(E_{\gamma})}}=\frac{\sigma^{CN}_{\gamma+^{61}Ni}(E_{\gamma})\Gamma^{^{61}Ni}_p (E_{ex})}{\sigma^{CN}_{\gamma+^{60}Ni}(E_{\gamma})\Gamma^{^{60}Ni}_p (E_{ex})}
\end{equation} 
\begin{figure}[h]
\includegraphics[scale=.6]{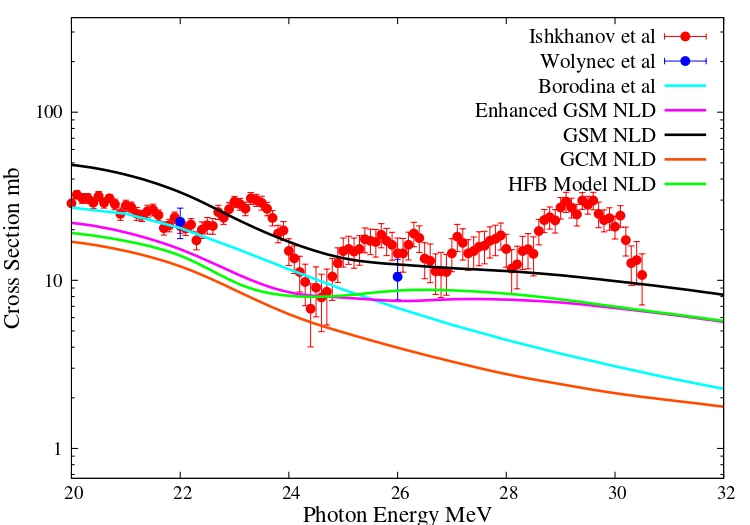}
\caption{\label{fig:four}Reference data of $^{60}$Ni($\gamma$,xp) reaction taken from EXFOR (Ishkhanov et al) compared with other data taken from EXFOR (Wolynec et al) and CINDA (Borodina et al) data libraries and EMPIRE 3.2.3 code.}
\end{figure}
Here $\sigma^{CN}_{\gamma+^{61}Ni}(E_{\gamma})$ and $\sigma^{CN}_{\gamma+^{60}Ni}(E_{\gamma})$ are the cross sections for the formation  of compound nucleus of desired and reference reactions. The photo-nuclear reaction $^{60}$Ni($\gamma$,xp) serves as the reference reaction and the related data is taken from corresponding EXFOR entry\citep{ishkhanov} as shown in Fig.\ref{fig:four} by red circles. Other data available in the libraries EXFOR\cite{wolynec} and CINDA\cite{borodina} are also shown in Fig.\ref{fig:four}. However, only two data points are available in the required energy range for Wolynec et al and the data of Borodina et al is evaluated one. The compound nuclear formation cross section of $^{61}$Ni and $^{60}$Ni were estimated using the nuclear reaction code EMPIRE 3.2.3 based on Hauser-Feshbach formalism, where GDR parameters for $\gamma$+$^{60}$Ni and $\gamma$+$^{61}$Ni are retrieved from RIPL-2. Utilising Eq.~(\ref{eq:two}) the cross section for $^{61}$Ni($\gamma$,xp) reaction have been calculated for excitation energy range 21 - 32 MeV, using the reference reaction cross section of $^{60}$Ni($\gamma$,xp), the experimental decay probabilities from $^{61}$Ni and $^{60}$Ni and the compound nuclear formation cross sections. 
The difference between the compound nuclear spin populated by the transfer reaction and the spin populated in the true photon induced reaction is found to be nearly 3$\hbar$. This is $< 10 \hbar$, which ensures the serendipitous matching condition for satisfying the surrogate ratio method\cite{schiba2010, schiba2011}. Hence it is attributed that the compound nucleus populated by the transfer reaction is really mimics the compound nucleus populated through the photon absorption.

\section{Nuclear Model Calculations}
The nuclear reaction code EMPIRE 3.2.3\cite{HERMAN} has been used for the theoretical calculations. Since proton is being tagged particle, the following channels have been accounted in the calculations viz, ($\gamma$,p), ($\gamma$,np), ($\gamma$,2p), ($\gamma$,pd) and ($\gamma$,p$\alpha$). Threshold energies for these reactions are tabulated in Table~\ref{Tab:2}. Theoretical calculation has been performed for energies ranging from threshold to 32 MeV. The cross section of reference reaction $^{60}$Ni($\gamma$,xp) measured by Ishkanov et al using bremsstrahlung photon obtained from betatron shows a number of well resolved maxima and minima \citep{ishkhanov}. They measured the cross section in steps of 100 KeV, from threshold up to 30 MeV. Hence the data has been optimized using theoretical calculations utilizing different level density models, optical models, gamma ray strength function models and GDR parameters.
 
\begin{table}[t]
\caption{Threshold energies for different reaction channels included in $^{61}$Ni($\gamma$,xp) reaction}
\begin{tabular}{c c  }

\hline\hline

Reaction Channel & \hspace{1mm} Threshold Energy (MeV) \\

\hline

$^{61}$Ni($\gamma$,p) &  9.8\\

$^{61}$Ni($\gamma$,np) & 17.35 \\

$^{61}$Ni($\gamma$,2p) & 18.14\\

$^{61}$Ni($\gamma$,p$\alpha$) & 17.02 \\

$^{61}$Ni($\gamma$,pd) & 22.49 \\

\hline\hline

\end{tabular}
\label{Tab:2}

\end{table}

\begin{figure}[b]
\includegraphics[scale= 0.6]{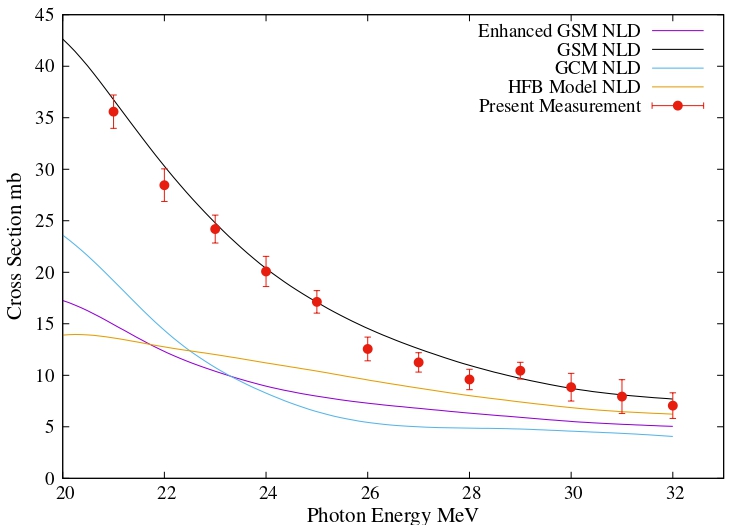}
\caption{\label{fig:f2}Comparison of $^{61}$Ni($\gamma$,xp) reaction data with different level density models obtained from nuclear reaction code EMPIRE 3.2.3.}
\end{figure}



\begin{figure}[h]
\includegraphics[scale=0.6]{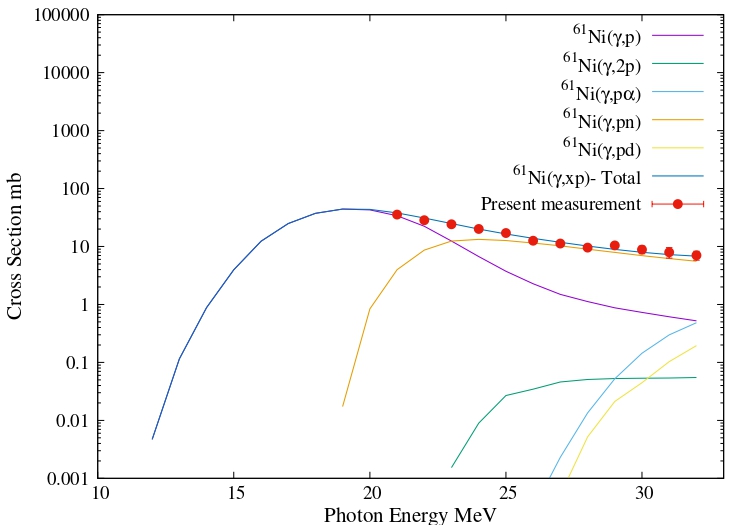}
\caption{\label{fig:f3} Cross section for different proton production channels of $^{61}$Ni($\gamma$,xp) reaction calculated using EMPIRE 3.2.3 code for GSM level density.}
\end{figure}

\paragraph*{} In the present study, the Hauser-Feshbach formalism\citep{PhysRev.87.366} is used for the compound nucleus models and pre-equilibrium part of the calculations has been done using exciton model\citep{KONING200415}. Necessary input parameters such as nuclear masses, nuclear level densities are taken from Reference Input Parameter Library RIPL-3\cite{ripl} and $\gamma$ ray strength functions, GDR parameters and discrete energy levels are taken from RIPL-2\cite{ripl2}. The phenomenological level density models such as Generalised super fluid model\citep{GSM},  Gilbert and Cameron model of nuclear level density\citep{Gilbert}, which incorporates constant temperature model for lower energy and Fermi gas model for higher energy region and parity-dependent nuclear level densities based on
the microscopic combinatorial model proposed by Hilaire and Goriely\citep{goriely} are applied for EMPIRE calculations. Along with this, Enhanced generalized super fluid model, that uses super fluid model below critical excitation energy and the Fermi gas model above critical excitation energy is also included in the calculations.
  Optical model potential due to Koning \& Delaroche\cite{Koning}, have been used for protons and neutrons, Avrigeanu et al\cite{Avrigeanu} are used for alphas, Haixia et al\cite{haixia} are used for deuteron. 
   Even though the energy range considered here is 20 - 32 MeV, Enhanced Generalized Lorentzian model, which is the sum of two Lorentzian is used for gamma ray strength function calculations. GDR parameters are chosen from the systematics based on the Dietrich and Berman\citep{DIETRICH1988199} compilation, which include experimental data for 150 nuclei ranging from $A=51$ to $A=239$. Mean free path parameters has a commenting role in determining pre-equilibrium contribution. In the present calculation mean free path parameter is taken as 1.5, so that the theoretical calculations are optimized to the present values. Considering $^{60}$Ni($\gamma$,xp) reference reaction data, it can be seen from Fig. \ref{fig:four} that the best fit is obtained for calculations using the generalized super fluid model (GSM) level density. \paragraph*{} The excitation function of $^{61}$Ni($\gamma$,xp) reaction, employing different level density models, obtained using theoretical code EMPIRE, along with measured data are plotted in Fig.~\ref{fig:f2}. Generalized Super fluid model of level density predicts the trend close to measurement. In the higher energy region all level density models show similar trend while in the lower energy region only GSM model is in good agreement with the result. Theoretically, the GSM model incorporates superfluid behaviour in the lower energy region and at higher excitation energy, fermi gas model is used to describe the system. Hence the corresponding paring correlations due to phase transition are also considered in the GSM level densities. The same model have been used in calculating the cross section for different channels viz, ($\gamma$,p), ($\gamma$,np), ($\gamma$,2p), ($\gamma$,pd)and ($\gamma$,p$\alpha$). These contributing channels, along with their sum, ($\gamma$,xp) at required photon energies are plotted in Fig.~\ref{fig:f3}. It is clear from the figure that the cross section for other proton emission channels such as $^{61}$Ni($\gamma$,np) and $^{61}$Ni($\gamma$,2p), are also prominent while compared to single proton emission cross section over the region of measured energies. This is because charged particle emissions are not strongly suppressed by the Coulomb barrier in the case of $^{61}$Ni isotope.
\begin{figure}[h]
\includegraphics[scale=0.6]{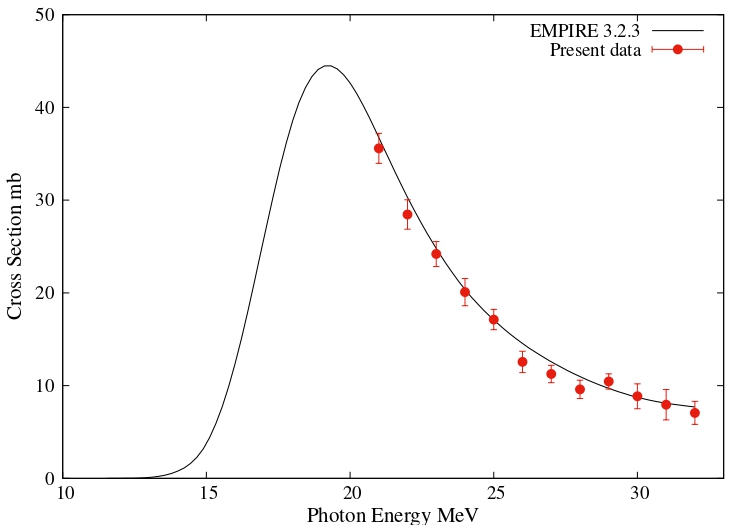}
\caption{\label{fig:f4} Total photo-proton cross section of $^{61}$Ni along with presently measured data}
\end{figure}

\section{Result and discussion}The cross section of the photo-nuclear reaction $^{61}$Ni($\gamma$,xp) is determined using the surrogate reaction method and results are analysed with different nuclear level density models obtained from statistical nuclear reaction code, EMPIRE 3.2.3. Measured cross sections are tabulated in Table \ref{Tab:3}. 
The reference reaction cross section was taken from EXFOR data library and optimized with theoretical model calculations. The same models have been used for the analysis of $^{61}$Ni($\gamma$,xp) surrogate data, in view of the fact that near by systems possess similar properties.
The cross section is determined over the incident photon energy range 21 - 32 MeV using the measured proton decay probabilities from the compound nuclei $^{61}$Ni$^{*}$ and $^{60}$Ni$^{*}$ . 
The result obtained is equivalent to the cross section for 100\% enrichment of $^{61}$Ni isotope.

\begin{table}[t]
\caption{Measured cross section for $^{61}$Ni($\gamma$,xp) reaction}
\begin{tabular}{c c }

\hline\hline

Energy MeV & \hspace{1mm} Cross section mb \\

\hline

21 & 35.584 $\pm$ 1.619\\

22& 28.452 $\pm$ 1.585\\

23 & 24.195 $\pm$ 1.353\\

24 & 20.079 $\pm$ 1.466\\

25 & 17.124 $\pm$ 1.094\\

26 & 12.558 $\pm$ 1.151 \\

27 & 11.252 $\pm$ 0.939\\

28 & 09.594 $\pm$ 0.988 \\

29 & 10.446 $\pm$ 0.817 \\

30 & 08.848 $\pm$ 1.345  \\

31 & 07.937 $\pm$ 1.639 \\

32 & 07.057 $\pm$ 1.243 \\

\hline\hline

\end{tabular}
\label{Tab:3}

\end{table}

\paragraph*{}
 In the theoretical calculations using EMPIRE, all possible photo-proton channels having threshold in the range 20 to 32 MeV are considered. The GDR cross section up to 30 MeV is given by Lorentzian shape with parameters characterising the total absorption of the giant dipole resonance. The excitation function obtained using theoretical code shows a broad peak centered at $\approx$ 20 MeV, which is shown in Fig.~\ref{fig:f3}, with GDR parameters $E_{1}=17.66$, $\sigma_{1}=69$, $\Gamma_{1}=6.26$, $E_{2}=19.85$, $\sigma_{2}=34.5$, $\Gamma_{2}=7.83$. Hence it is seen from the figure that the experimental cross section obtained using surrogate method is at the tale end of GDR region.

\section{Summary and Conclusion}
In summary, we have determined the cross section of $^{61}$Ni($\gamma$,xp) by employing surrogate ratio method. The cross section obtained is in steps of 1 MeV in the equivalent photon energy range 21 - 30 MeV. Theoretical model calculations performed using Empire 3.2.3 code are in consistent with the measured results. Presently, surrogate-reaction method successfully produce the cross section of photonuclear reaction $^{61}$Ni($\gamma$,xp). The isotope we considered here is $^{61}$Ni, which is a medium mass nuclei. In the case of medium mass nuclei, where giant dipole resonance is found to be the dominant mechanism up to 30 MeV, photo-proton and photo-neutron cross sections are more or less comparable. Hence, the proton decay probabilities can be studied by populating required compound nucleus via surrogate technique, which can further used to determine the corresponding cross sections. Accordingly, this indirect method can be used as an effective tool to study the cross sections of photonuclear reactions involving medium mass nuclei.
\section{Acknowledgments}
Authors acknowledge DAE-BRNS for the partial financial support under the project No. 36(6)/14/30/2017-BRNS/36204. The authors also acknowledge Department of Science and Technology under the project No. YSS/2015/001842 for the partial financial support during the experiment. We are thankful
to the Pelletron-LINAC staff for providing beam during the experiment.
\bibliography{reference}
\end{document}